\pdfoutput=1
\documentclass[DTMColor]{vdpreprint}
\usepackage{graphicx}
\usepackage{xcolor}
\hypersetup{pdftitle={Managing Verification Debt: A Framework for Responsible Human-AI Research Partnerships},pdfauthor={Abdulkadir Celik and Bashir M. Al-Hashimi},pdfkeywords={human-AI partnership; verification debt; temporal asymmetry; research supervision; provenance}}

\begin{document}

\runningtitle{Managing Verification Debt}
\authormark{Celik and Al-Hashimi}

\articletype{COMMENTARY}

\title{Managing Verification Debt: A Framework for Responsible Human--AI Research Partnerships}

\author[1]{Abdulkadir Celik\hyperlink{corr}{*}}
\address[1]{School of Electronics and Computer Science, University of Southampton, Southampton, SO17 1BJ, UK}

\author[2]{Sir Bashir M. Al-Hashimi}
\address[2]{King's Institute for AI, King's College London, London, WC2R 2LS UK}
\address{\hypertarget{corr}{*}Corresponding author. \email{a.celik@soton.ac.uk}}

\keywords{human--AI partnership; discipline; temporal asymmetry; verification debt; research supervision; provenance}

\abstract{
Conversation with generative artificial intelligence (AI) can quickly turn a tentative scientific question into the outline of a research programme. Although this capability can accelerate research, it also places two closely related pressures on human intelligence: 1) \textit{temporal asymmetry} between rapid AI generation and slower expert evaluation, and 2) a \textit{cognitive gap} between the available effort, expertise, attention and capacity to assess AI outputs. When unexamined AI contributions are carried into subsequent claims or decisions, these pressures produce \textit{verification debt}: validation work that remains outstanding before a contribution can be relied upon. If left unresolved, this debt can undermine responsible AI in research, understood here as the transparent and traceable use of AI under effective human oversight and accountability. This commentary argues that responsible human--AI research partnerships therefore require a disciplined method for managing verification debt, not better prompting alone. We propose a shared topology of the research programme: a structured representation that connects questions, claims, assumptions, sources, tests and decisions while recording their provenance, dependencies and verification status. The topology makes unresolved debt visible, enables corrections to propagate through dependent claims and requires consequential contributions to be examined at decision gates before research proceeds. We consider its application in two-way partnerships and supervised partnerships that extend from doctoral research to multi-way research teams, and examine its implications for scholarly assessment, peer review and empirical evaluation.
}

\maketitle

\section{Introduction}\label{sec:intro}
Scientific research increasingly unfolds through conversations with generative AI. A researcher may begin with a tentative question, yet within a dozen prompts a large language model (LLM) can identify literature, propose mechanisms, outline experiments, develop code, interpret preliminary results and generate counterarguments~\cite{wang2023}. What began as a single question can therefore acquire the appearance of a coherent research programme before its central claims have been verified or adequately supported by evidence. Although this capacity can broaden the range of questions researchers explore and accelerate demanding work, the resulting coherence can contribute to an \emph{illusion of understanding}, in which researchers feel that they understand more deeply than the available evidence warrants~\cite{messeri2024}.

Responsible AI in science must therefore be realised through a trustworthy human--AI research partnership design, which requires transparency about AI contributions, traceability of claims and evidence, human oversight of consequential decisions and clear accountability for scholarly contributions~\cite{unesco2023}. Accordingly, a responsible partnership must address not only what AI can generate and how quickly it can do so, but also what researchers can understand, verify and take responsibility for. 

In human--AI partnership, we observe two closely related pressures on human intelligence: 1) \emph{temporal asymmetry} describes the difference between the speed of AI generation and the time required for expert evaluation, and 2) the \emph{cognitive gap} describes the difference between the intellectual effort required to evaluate those outputs and the expertise, attention and cognitive capacity available for that evaluation. Cognitive load contributes to this gap, especially when models produce extensive, interconnected material outside a researcher's immediate expertise~\cite{sweller1988}.

When AI-generated contributions are carried into later claims or decisions before adequate examination, these temporal and cognitive pressures produce \emph{verification debt}: the accumulated validation work that must be completed before a claim, method or result can be relied upon~\cite{zentner2026}. Verification debt does not imply that an output is false; it rather signals that the basis for relying on it remains incomplete. As the debt grows, valuable ideas may be overlooked, errors and unsupported conclusions may propagate, scientific judgement may become diluted and researchers may become dependent on outputs they have insufficient capacity to verify~\cite{messeri2024}.

Since managing this debt requires more than better prompting, we propose a \emph{shared topology of the research programme}: a structured representation connecting the questions, claims, assumptions, sources, tests and decisions that constitute a research programme. The topology records where consequential ideas/contributions originated, how they depend on one another, whether they have been checked, by whom and when. Corrections can therefore be traced through dependent claims. Before the research proceeds through a decision gate, any remaining debt must be resolved or explicitly accepted.

Building on these concepts, this commentary develops a topological discipline for responsible human--AI research partnerships. We consider how the topology can be applied in two practical settings: two-way partnerships between an independent researcher and AI agents, and supervised partnerships that extend from doctoral research to multi-way research teams. We then examine its implications for scholarly assessment, peer review and empirical evaluation. Our central proposition is that the future of scientific discovery depends not only on accelerating AI generation, but also on strengthening the human capacity for verification.

\section{The Temporal and Cognitive Origins of Verification Debt}\label{sec:asymmetry}

The following subsections examine how the two pressures introduced above operate in research practice, how they reinforce one another and when deferred evaluation becomes a scientific liability.

\subsection{Temporal asymmetry: generation outpaces evaluation}

AI may produce an explanation, literature review, derivation, simulation or data analysis within moments. Establishing whether that contribution is reliable may require reading the original sources, examining assumptions, reproducing a result, consulting another expert or conducting an experiment. These activities remain time-consuming even when the AI-generated contribution ultimately proves sound.

The asymmetry cannot be inferred from the length of an output or treated as a fixed difference between machine and human speed. A brief response may introduce a consequential assumption that requires days of investigation, whereas a longer synthesis may be straightforward to assess when it concerns familiar material. Hence, the time required for evaluation depends on the novelty of the contribution, the available evidence, the researcher’s expertise and the consequences of relying upon it. Although more capable models can assist with verification, they may also produce greater volumes of interconnected material that researchers must assess independently or carry the work into fields beyond their immediate expertise. AI can therefore accelerate both the production of potentially valuable contributions and the evaluation burden they create.

\subsection{The cognitive gap: output exceeds available capacity}

Evaluation demands more than time. Researchers must understand how an AI-generated contribution relates to the question, which assumptions it introduces, what evidence supports it and how an error would affect the rest of the work. The required effort may exceed the expertise, attention and cognitive capacity available, particularly when the outputs are extensive or interdependent.

Cognitive load contributes to this gap when an interaction produces more information and complex relationships than a researcher can hold and examine at once~\cite{sweller1988}. The problem becomes significantly pronounced when the material lies outside the researcher's field of expertise or combines several disciplines. A specialist may quickly recognise a familiar limitation, whereas a doctoral researcher or a collaborator from another field may need substantial background work before reaching the same judgement. Even an experienced researcher can lose track of which statements came from evidence, which originated as AI suggestions and which were introduced merely to sustain the human--AI conversation.

Temporal asymmetry and the cognitive gap consequently reinforce one another. Rapid generation increases the volume and interdependencies of the material requiring attention, while limited time makes it harder to construct the comprehension necessary for careful evaluation. Deferred examination can widen the gap further because sources, assumptions and earlier decisions become more difficult to reconstruct. The resulting challenge is not evidence of an individual's inadequacy; it is an unavoidable consequence of combining machine-scale generation with bounded human attention and domain-specific expertise. It also helps explain why combining capable humans and AI systems does not automatically improve performance~\cite{vaccaro2024}. The outcome therefore depends on how the partnership is designed to allocate responsibility and ensure that human and AI contributions are systematically organised, examined and combined.

\subsection{When verification debt begins to accumulate}

Temporal asymmetry and the cognitive gap can postpone evaluation, but delay alone does not create verification debt. The scientific liability arises only when an unresolved contribution is carried into a subsequent claim or decision.

A speculative idea may therefore remain unverified during exploration without creating a scientific liability, as long as its provisional status is visible and it is not treated as established evidence. Once that idea supports an experimental design, informs an analysis or becomes a premise for another claim before it has received adequate scrutiny, verification debt begins to accumulate. Each dependent step raises the consequence of an unresolved error and makes the original basis more difficult to reconstruct. Familiarity with the resulting line of reasoning can then be mistaken for understanding, contributing to the \emph{illusion of understanding} described in AI-assisted science~\cite{messeri2024}. 

The latest AI-generated mathematics illustrates how this process can extend beyond an individual conversation. OpenAI's claimed result on the Navier--Stokes Millennium Prize Problem emerged from a largely autonomous multi-agent search in which human researchers set objectives, directed resources and consolidated intermediate findings~\cite{openai2026ns}. Even though formal verification can establish that a proof follows from an encoded statement, human scrutiny remains necessary to determine whether that statement faithfully represents the intended mathematical problem and what the result contributes conceptually. Mathematicians were still digesting the lengthy manuscript when, one month later, OpenAI released more than 700 AI-generated mathematical manuscripts across 372 problem families~\cite{openai2026math}. The results were recorded at different stages of verification, and some entries were subsequently revised or withdrawn. At this scale, production advances faster than the scientific community can understand, audit and assign responsibility for what has been produced.

The debt metaphor nevertheless has limits. Scientific research never reaches a state in which every proposition has been established beyond doubt as attempting exhaustive verification would bring progress to a halt. The practical objective is to identify which unresolved contributions matter for the next consequential decision. We refer to these critical points as \emph{decision gates}. Before a gate is passed, the claims on which the next stage depends should be examined, and any remaining debt should be either resolved or explicitly accepted by an accountable human decision maker. Exploration can remain open between gates, while unverified contributions are prevented from silently acquiring the status of evidence. Section~\ref{sec:topology} develops the practical framework through which this debt can be made visible and managed.

\section{A shared topology for disciplined management of verification debt}\label{sec:topology}

Managing verification debt requires more than retaining a transcript of the human--AI interaction. Researchers need to see how a question develops into claims, which evidence supports those claims, what assumptions remain unresolved and which outcomes depend upon them. We define the \emph{topology} as a shared, structured map of how the research programme develops and is verified. It connects questions, claims, sources, assumptions, tests and outcomes through defined relations, while showing their origin, verification status, timestamps and decision gates. The topology represents the contributions and dependencies made explicit by the human--AI partnership rather than the LLM's internal reasoning.

\begin{figure}[t]
\centering
\includegraphics[width=\textwidth]{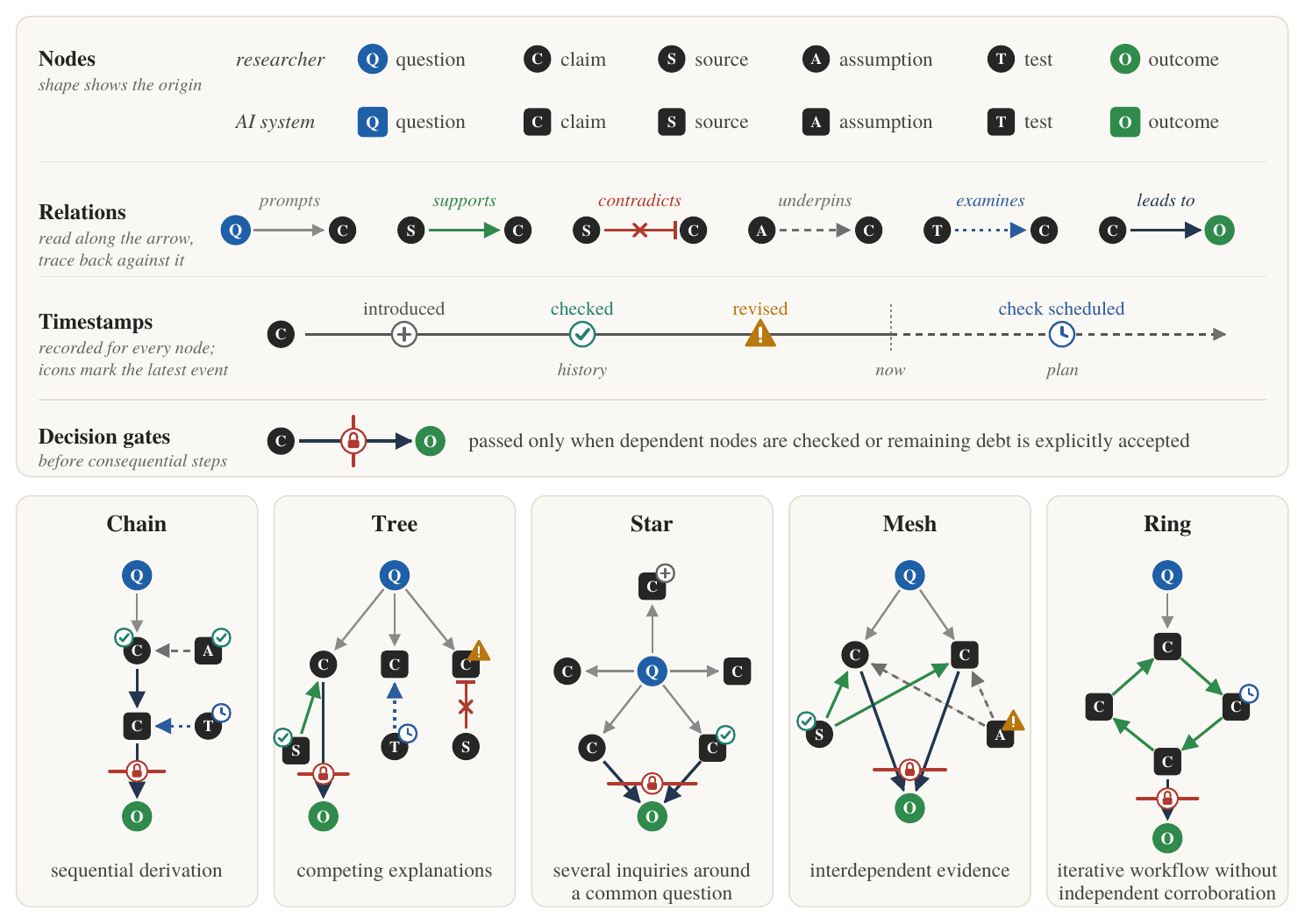}
\caption{The visual vocabulary of the topology (upper panel) and five illustrative dependency patterns through which a research programme may develop (lower panels)}
\label{fig:topology}
\end{figure}

The upper panel of Figure~\ref{fig:topology} introduces the visual vocabulary used to construct the topology. Its six node types represent the principal elements that researchers may need to trace from the beginning of a research programme to its outcomes. A \emph{question} (Q) defines what the research seeks to investigate; a \emph{claim} (C) proposes an answer or explanation; a \emph{source} (S) provides external evidence; an \emph{assumption} (A) states a condition on which a claim depends; a \emph{test} (T) examines whether the claim holds; and an \emph{outcome} (O) records what can be concluded from the preceding claims and evidence. These six types are not intended to capture every element of scientific reasoning but to identify those most important for tracing verification. The shape of each node indicates whether the contribution originated from a researcher (circle) or an AI system (square), while the connecting arrows show how the nodes relate to one another.

A question may \emph{prompt} a claim; a source may \emph{support} or \emph{contradict} it; an assumption may \emph{underpin} it; a test may \emph{examine} it; and a sufficiently supported claim may \emph{lead to} an outcome. Timestamps show when each node was introduced, checked or revised, as well as when a future check is scheduled. The final row introduces the \emph{decision gate}, which may be passed only after the dependent nodes have been checked or any remaining verification debt has been explicitly accepted by an accountable human decision maker. Each critical node should also name its responsible reviewer and the gate at which verification is required. Together, the nodes and their connections record the origin, evidential basis and verification status of each contribution; the topology makes clear what must be verified before the research proceeds.

This separation of origin from evidential support is important. A claim proposed by a researcher is not necessarily reliable, while an AI-generated claim may become reliable after independent examination. Similarly, attaching a paper to a claim does not establish support until the paper has been read and shown to provide the required evidence. Earlier work on computer-supported argumentation and scholarly knowledge representation has modelled claims and their relations~\cite{scheuer2010}, while the PROV ontology provides a standard vocabulary for provenance~\cite{prov2013}. The proposed topology builds on these ideas by connecting provenance and verification to the decisions that shape an active research programme.

The lower panels of Figure~\ref{fig:topology} illustrate five dependency patterns through which a research programme may develop: chain, tree, star, mesh and ring. The pattern need not be prescribed in advance; it reflects the research question and may emerge or change naturally as researchers and AI systems develop the programme together. The chain provides the simplest illustration. In the example shown, a question prompts an initial researcher-origin claim that rests on an AI-origin assumption. Both nodes have been checked before the first claim leads to a second claim proposed by the AI system. A researcher-origin test of that second claim remains scheduled, so the decision gate stays closed and the outcome cannot yet be adopted. Relying on the second claim before completing the test would carry verification debt across the gate. If the test contradicts the claim or the underlying assumption is later revised, the dependent claim and outcome must return to review. This propagation is comparatively easy to trace in a chain because the research programme follows a largely sequential path.

The other four patterns illustrate how the same principle operates as a research programme becomes less linear. A \emph{tree} arises when a question develops into competing explanations or alternative methods, so an unresolved contribution may affect one branch or several branches that share the same basis. A \emph{star} organises several lines of investigation around a common question, allowing each line to be examined separately while retaining its connection to the central problem. A \emph{mesh} contains interdependent claims and evidence, which means that revising one assumption may require several connected claims to be reconsidered. A \emph{ring} represents an iterative workflow in which claims repeatedly inform one another; it requires particular caution because circulation within a closed argument does not provide independent corroboration.

These patterns are illustrative rather than exhaustive or mutually exclusive. A human--AI interaction may begin as a chain, develop into a tree as alternative explanations emerge and later form a mesh as evidence becomes shared across several claims. Multiple patterns may therefore coexist within the same project. Verification debt can arise in any pattern, but the effort required to locate and resolve it generally increases as claims acquire more dependencies and corrections can propagate along multiple paths. 

Construction and maintenance of the topology need not be entirely manual. AI agents could extract candidate nodes and relations from research conversations, link claims to sources and tests, record timestamps and flag dependencies affected by a revision. Such automation would reduce the documentation burden, but it cannot confer verification: human experts must validate the topology, correct omissions or misclassified relations, determine verification status and retain control of decision gates. The topology may therefore be generated and updated with AI assistance, while its scientific meaning and governance remain human responsibilities.

\section{Exercising the management discipline across research partnerships}\label{sec:partnerships}

\begin{figure}[t]
\centering
\includegraphics[width=\textwidth]{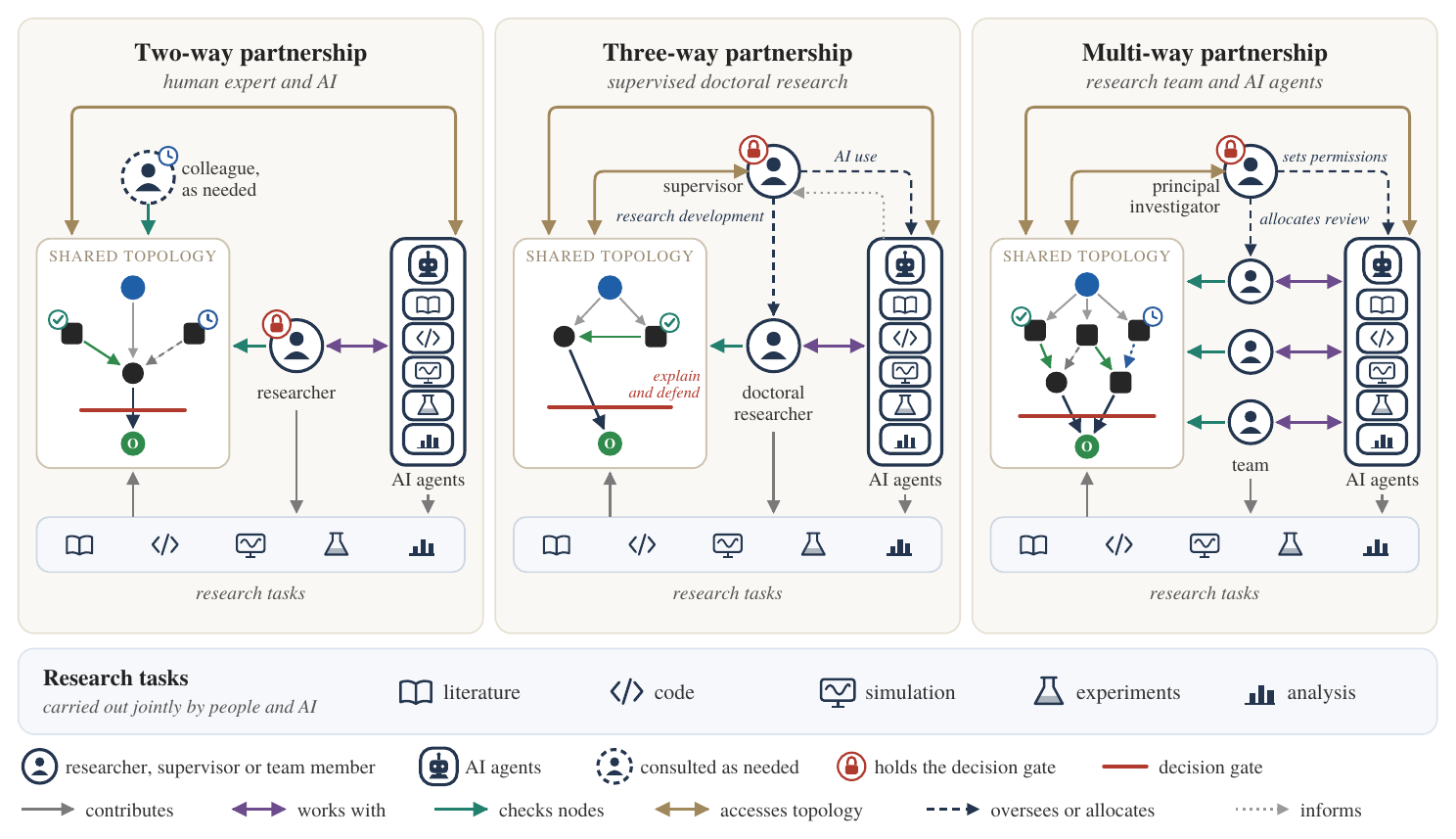}
\caption{Two-way, three-way and multi-way human--AI research partnerships around a shared topology}
\label{fig:partnerships}
\end{figure}

How the topology is maintained, reviewed and governed depends on the design of the human--AI partnership. Temporal asymmetry and cognitive gap depend not only on the task but also on who must verify the output: a claim that an expert can assess in minutes may take a novice days or lie beyond the competence of a researcher from another field. Across all partnership arrangements in Figure~\ref{fig:partnerships}, researchers and AI agents may contribute to literature review, code, simulations, experiments and analysis. AI agents can also maintain source links and distinguish provisional suggestions from checked contributions. Human participants nevertheless frame the research question, judge the adequacy of evidence and retain responsibility for consequential decisions.

The person formally accountable for the work---an independent researcher, supervisor or principal investigator---sets the agents' permissions, assigns responsibility for verification and controls the decision gates. The topology supports this role by recording the origin and status of consequential contributions, identifying who must review them and keeping unresolved debt visible. What changes across the arrangements in Figure~\ref{fig:partnerships} is the distribution of human expertise, the process through which verification is conducted and the opportunity for that expertise to develop.

\subsection{Two-way partnership: human expert and AI collaboration}

A two-way partnership joins an independent researcher with one or more AI agents in a shared programme of work (Figure~\ref{fig:partnerships}, left). The researcher holds the accountable role and must provide context, recognise consequential assumptions and decide when an AI-generated contribution requires further expertise or experimental testing. The topology supports these decisions by separating provisional outputs from contributions that have received an adequate evidential basis.

This judgement remains bounded by domain expertise. A senior researcher may readily assess an output concerning a familiar method while lacking the knowledge needed to evaluate a claim from another field. In such cases, consequential assumptions become harder to recognise and take longer to examine. A node that lies outside the researcher's expertise should therefore be assigned to an appropriate reviewer, while the associated gate remains closed until that review is complete.

The partnership should also strengthen rather than displace the researcher's judgement. Greater confidence in generative AI was associated with lower self-reported critical-thinking effort in a survey of 319 knowledge workers~\cite{lee2025}. Although this finding does not establish a lasting loss of skill, it raises an important criterion for partnership design: AI assistance should leave the researcher better able to understand, defend and revise the work, rather than merely enabling the rapid production of a plausible result.

\subsection{From three-way doctoral training to multi-way research teams}

A three-way partnership places a doctoral researcher, supervisor and AI agents within the same research and training environment (Figure~\ref{fig:partnerships}, centre)~\cite{alhashimi2026}. The supervisor oversees both the student's intellectual development and the AI-generated contributions entering the research programme. The student remains responsible for understanding and defending the work, while the role of AI extends beyond performing delegated research tasks. AI agents may identify missing background knowledge, adapt explanations to the student's level, compare the student's reasoning with relevant methods and sources, and suggest questions or exercises that expose unresolved assumptions. Used in this way, AI can help the supervisor provide more responsive training alongside research guidance.

These capabilities do not amount to educational judgement. An apparent gap in the student's explanation may reflect incomplete knowledge, unfamiliar terminology, a poorly specified task or a defensible disagreement. The supervisor must interpret such indications in context and determine the appropriate learning response. AI should therefore augment the supervisor's capacity to develop the student's expertise and judgement rather than replace the intellectual relationship between supervisor and student. The supervisor must also distinguish successful AI-assisted output production from the student's independent understanding and capacity to exercise judgement~\cite{mit2026}.

The topology gives this combined research and training responsibility a practical form. A student might prepare an initial synthesis, compare it with an AI-assisted account and then defend the representation of sources, the proposed research gap and a consequential methodological assumption. The topology allows the supervisor to see not only which claims require verification but also where the student may need further training before evaluating those claims independently. Supervision meetings can then concentrate on these dependencies rather than reconstructing the entire AI conversation. Over time, the topology may reveal the development of independent judgement while helping the supervisor identify persistent training needs. The supervisor must nevertheless decide which parts of the topology should be retained, who may inspect them and how the student's contribution should be assessed~\cite{unesco2023}.

Larger and more demanding projects may extend this three-way arrangement into a multi-way partnership involving postdoctoral researchers, technical experts, doctoral students, and multiple specialised AI agents (Figure~\ref{fig:partnerships}, right). The supervisor's accountable role then develops into that of a principal investigator or designated research lead, while verification is distributed among team members according to expertise. 

The topology scales with this extension by recording who established each objective, which person or agent generated each consequential contribution, what tools or model versions were used and who examined the result. Agreement among several agents does not constitute independent evidence when they rely on the same sources, methods or previous outputs; without scrutiny of those dependencies, repeated circulation may reproduce the ring pattern in Figure~\ref{fig:topology} without corroborating the claim. The principal investigator therefore retains control of consequential decision gates, while the strictness of those gates should reflect the task, the available means of verification and the consequences of error.

Taken together, the topology and partnership design translate core responsible-AI principles into research practice. Distinguishing human- and AI-origin contributions supports transparency, while linking claims to sources, assumptions, tests and revisions provides traceability. Decision gates preserve human oversight, while named reviewers and accountable research leads clarify who accepts or resolves verification debt. The framework does not by itself guarantee fairness, privacy or safety, but it makes the dependencies and responsibilities through which such concerns must be addressed visible.

\section{From Human–AI Partnership to Responsible Scholarship}\label{sec:assessment}

The responsible-AI principles established within the partnership must remain visible when research leaves it in the form of a thesis, manuscript or proposal. Otherwise, unresolved verification debt is transferred to examiners and reviewers who lack the time and context available to the original researchers.

\subsection{Assessing scholarly value and independent judgement}

Generative AI further weakens the already imperfect relationship between the volume of scholarly output and the value of a contribution. When drafting, coding and analysis can be accelerated, publication counts reveal even less about the quality of the research question, the strength of its evidence or the researcher's command of the work. Recruitment, evaluation and promotion should therefore place greater weight on methodological care, well-supported findings and contributions that others can examine and extend. DORA provides an established basis for evaluating research on its substance and recognising a wider range of scholarly contributions~\cite{dora2013}.

The same principle applies to doctoral assessment. A polished thesis cannot by itself demonstrate originality, expertise or intellectual independence. During the viva or defence, the candidate must explain the research, defend consequential decisions and respond to academic challenge. A thesis accompanied by its topology would make consequential AI contributions transparent and allow examiners to trace how they were verified. It would also provide a structured basis for asking why a decision gate was passed and how a conclusion would change if a supporting node were revised. Such examination assesses the independent judgement that the supervised partnership in Section~\ref{sec:partnerships} is intended to develop.

\subsection{Verification debt in peer review}

Verification debt does not disappear when research is submitted; unless addressed by the authors, it is transferred to reviewers. Publication volume has long grown faster than the pool of scientists available to review it, a trend that generative AI intensifies because manuscripts and proposals can be produced at machine speed, whereas their evaluation remains constrained by expert time. Reviewers cannot realistically reconstruct every AI-assisted analysis, recover every source or identify every assumption on which a conclusion depends.

 Reviewers are also increasingly turning to AI to draft or refine their assessments. AI assistance may help organise material or identify questions, but repeated model agreement does not constitute independent evidence. When AI-generated claims are assessed through AI-generated reviews without sufficient expert scrutiny, the ring pattern in Figure~\ref{fig:topology} reappears at the level of the scholarly system: circulation begins to substitute for verification. Sharing a compact topology would instead give reviewers structured access to the dependencies and unresolved debt that matter most for their evaluation.

\subsection{The topology for transparent accountability}

A compact topology could accompany selected manuscripts, proposals and theses as a structured practice of transparency and traceability. It should identify the principal claims and dependencies, mark consequential AI-generated contributions, record who examined them and show which verification debt was resolved or explicitly accepted at each decision gate. Examiners and reviewers could then direct their limited attention towards consequential dependencies rather than reconstructing the research programme from the final narrative alone.

Transparency does not require disclosure of every prompt or intermediate exchange. Access to the topology should remain proportionate to the purpose of assessment and protect confidential data, unpublished ideas and personal information. The topology must nevertheless contain enough information for another researcher to recover and scrutinise the basis of the principal conclusions.

The topology would complement existing forms of disclosure rather than replace them. Contributor-role taxonomies describe human contributions, while many journal policies require disclosure of relevant AI use and do not recognise AI tools as authors. Such statements identify whether AI was used, whereas the topology would show how consequential AI-generated contributions were examined and who accepted responsibility for the resulting decisions. Authorship therefore continues to carry human accountability, while the topology makes the exercise of human oversight more specific and examinable. Its practical value should be judged by whether it improves scrutiny without imposing a documentation burden disproportionate to its scientific benefit.

\section{Concluding remarks and evaluation priorities}\label{sec:evaluate}

Because the proposed discipline remains a set of testable propositions about research practice, its benefits should be demonstrated rather than assumed. Studies could compare ordinary AI-assisted work with work organised through a shared topology across the partnership arrangements described in Section~\ref{sec:partnerships}. Evaluation should address three classes of outcome:
\begin{itemize}
\item \textbf{Verification quality} concerns the recovery of sources and assumptions, detection of consequential errors and propagation of corrections through dependent claims.
\item \textbf{Decision quality} concerns whether consequential gates are passed on an adequate evidential basis. The topology can operationalise, rather than directly measure, verification debt through indicators such as the number, age, consequence and dependency depth of unchecked nodes, together with the time required to resolve them.
\item \textbf{Human development and cost} form the third class of outcome. Researchers should be able to explain, apply and revise their work after AI assistance has been removed, while evaluation should distinguish domain expertise from career seniority and examine whether AI agents help supervisors identify genuine training needs. Studies should also record the time required to maintain the topology, the workload placed on supervisors and principal investigators, and any effect on the quality or efficiency of peer review.
\end{itemize}

These potential benefits must be assessed against the topology's own risks. A polished diagram may give a weak argument unwarranted authority, while a recorded check may conceal superficial examination. The topology must therefore be judged against its sources, tests and expert scrutiny rather than accepted as evidence in itself. Verification debt should therefore become an integral consideration in research practice for responsible human--AI partnership. This commentary has provided the rationale for doing so and a structured approach that makes unresolved verification visible, assigns responsibility and prevents consequential decisions from advancing without examination. The framework should now be tested and refined across disciplines. We hope it provides a durable basis for using AI in scientific discovery without weakening the human understanding, judgement and accountability on which science depends.
\begin{con}
\ctitle{Author Contributions}
A.C. and S.B.M.A.-H. conceived the commentary. A.C. developed the verification-debt framework and the research topology, designed the figures and wrote the manuscript. S.B.M.A.-H. contributed to the framing of human--AI research partnerships and critically reviewed and revised the manuscript. Both authors approved the final version.

\ctitle{Financial Support}
This research received no specific grant from any funding agency, commercial or not-for-profit sectors.

\ctitle{Conflicts of Interest}
The authors declare no conflicts of interest exist.

\ctitle{Ethical Approval}
Not applicable. This commentary reports no original study involving human participants or animals.

\ctitle{Acknowledgements}
OpenAI ChatGPT and Anthropic Claude were used to assist with language editing. Claude was also used to render the figures as vector graphics from the manual designs sketched by the authors. All authors reviewed the content and take full responsibility for the paper.
\end{con}

\label{LastPage}
\end{document}